\documentclass[aps,preprint,groupedaddress,nofootinbib]{revtex4-1}
\usepackage{float}
\usepackage{amssymb}
\usepackage{amsmath}
\usepackage{graphicx}
\usepackage{stackrel}
\usepackage{natbib}

\begin{document}
	\title{Intensification of Gravitational Wave Field Near Compact Star}
	
	\author{Ashadul Halder}
	\email{ashadul.halder@gmail.com}
	\affiliation{Department of Physics, St. Xavier's College, \\30, Mother Teresa Sarani, Kolkata-700016, India.}
	
	\author{Shibaji Banerjee}
	\email{shiva@sxccal.edu}
	\affiliation{Department of Physics, St. Xavier's College, \\30, Mother Teresa Sarani, Kolkata-700016, India.}
	
	\author{Debasish Majumdar}
	\email{debasish.majumdar@saha.ac.in}
	\affiliation{Astroparticle Physics and Cosmology Division, Saha Institute of Nuclear Physics, HBNI\\1/AF Bidhannagar, Kolkata-700064, India.}
	
	\date{}
	\begin{abstract}
		The gravitational waves (GWs) has been a topic of interest for its versatile capabilities of probing several aspects of cosmology and early Universe. Gravitational lensing enhances further the extent of this sort of waves and upgrade our understanding to a next level. Besides several similarities with optical waves, GWs are capable of passing through optically opaque celestial objects like stars, exoplanets unlike light waves and manifest a different kind of lensing effect. In this work we have explored the lensing action of compact objects on gravitational waves using numerical means. After modeling the internal mass distribution of the compact objects by TOV equations and tracing wavefronts using geodesic equations, we have found that the GWs are indeed lensed in a manner analogous to the optical lensing of light in presence of a thick optical lens by producing spherical aberration in the focused waves. The distance to the best focused point shows significant dependence with the mass and radius of the lensing star and unlike gravitational lensing, the region inside and outside compact objects responds differently to the incoming waves.  
		
	\end{abstract}
	\maketitle

		The gravitational waves (GWs) have gained immense importance after first detection \cite{GW150914} for their versatile capabilities of probing several aspects of cosmology and early Universe. These newly discovered waves have been generated by some of the most violent cosmological events such as black hole - black hole mergers \cite{GW150914}, neutron star merger \cite{GW170817} etc. and carry signatures of them. Within the framework of general theory of relativity (GR), gravitational waves are in general disturbances of space-time that have many semblances with the propagation of electromagnetic waves. 
		Beyond these similarities, the gravitational waves are also bestowed with some additional properties which make them a significant observational tool \cite{unscathed_1,unscathed_2,unscathed_3}.

		Strong gravitational lensing \cite{schutz,strong_lens_1,strong_lens_2} have turned out to be a most relevant tool in several crucial cases of astrophysical measurements \cite{hubble_const_lens} viz. measuring the masses of black holes \cite{bh_lens_1,bh_lens_2}, estimation of density profile of galaxies \cite{galaxy_lens_1,galaxy_lens_2} etc.
		Being light-like \cite{Hobson} in nature, GWs propagate along null geodesics \cite{Hobson} and produce lensing phenomenon similar to the electromagnetic waves (EM waves). 
		Unlike the EM waves GWs, which are ripples of space-time, are capable of passing through optically opaque celestial objects like stars, exoplanets and manifest a different kind of lensing effect. 
		In the event of gravitational wave encounters such an object in the course of their propagation, the GWs suffer an additional modifications of different nature in the star and change its form. Therefore, gravitational waves may also be a tool for obtaining useful information about the deep interior of compact object.

		A consequence of GR is that the presence of mass distorts the space-time fabric. We started with the most simplest form of a metric that describes the space-time curvature due to a massive object is Schwarzschild \cite{Hobson,schutz} metric given by,
		
		\begin{equation}
			c^{2}d\tau^{2}=c^{2}\left(1-\frac{2GM}{c^{2}r}\right)dt^{2}-\left(1-\frac{2GM}{c^{2}r}\right)^{-1}dr^{2}-r^{2}(d\theta^{2}+\sin^{2}\theta d\varphi^{2}),\label{eq:line element schwarzschild}
		\end{equation}
		where, $\tau$ denotes the proper time, $(r,\theta,\varphi)$ are the coordinates of spherical coordinate system, $M$ is the mass, which is responsible for the curvature in space time and the other notations used have their usual significance.

		Since gravitational waves propagate with the speed of light, the proper time $\tau$ would vanish \cite{schutz,Hobson} in this case. Now, with $d\tau=0$ in Eq.~\ref{eq:line element schwarzschild}, the equation of null-geodesics in two dimensional plane can be obtained with time $t$ as the path parameter, given by
		\begin{equation}
			\frac{d^{2}r}{dt^{2}}=\frac{2GM}{c^{2}r^{2}-2GMr}\left(\frac{dr}{dt}\right)^{2}+\left(r-\frac{3GM}{c^{2}}\right)\left(\frac{d\theta}{dt}\right)^{2},\label{eq: r_eq}
		\end{equation}
		\begin{equation}
			\frac{d^{2}\theta}{dt^{2}}=\frac{2(3GM-c^{2}r)}{r(c^{2}r-2GM)}\frac{dr}{dt}\frac{d\theta}{dt}.\label{eq:theta eq}
		\end{equation}

		These two equations (Eqs.~\ref{eq: r_eq}, \ref{eq:theta eq}) describe the trajectories of gravitational waves during their propagation. It is to be mentioned that, mass $M$ in Eqs.~(\ref{eq: r_eq}) and (\ref{eq:theta eq}) is constant outside the lensing star. However the numerical value of the effective mass should change with radial vector ($r$) inside the star. The variation of mass $M$ inside a compact massive stars such as a neutron star or a quark star \cite{Witten_qs} can be evaluated by solving Tolman-Oppenheimer-Volkov equations (TOV) \cite{TOV} numerically. The TOV equation is given by,

		\begin{equation}
			\frac{dP}{dr}=-\frac{GM}{r^{2}}\rho\left(1+\frac{P}{\rho c^{2}}\right)\left(1+\frac{4\pi r^{3}P}{Mc^{2}}\right)\left(1-\frac{2GM}{c^{2}r}\right)^{-1}.\label{eq:TOV}
		\end{equation}

\begin{figure}
			\centering{}
			\caption{\label{fig:profile_of_ns}Density profile (a) and corresponding mass profile (b) of a neutron star parameterized by central density $2.5\times 10^{17}\:\rm{kg/m}^{3}$.}
			\begin{tabular}{cc}
				\includegraphics[width=7cm,height=5cm]{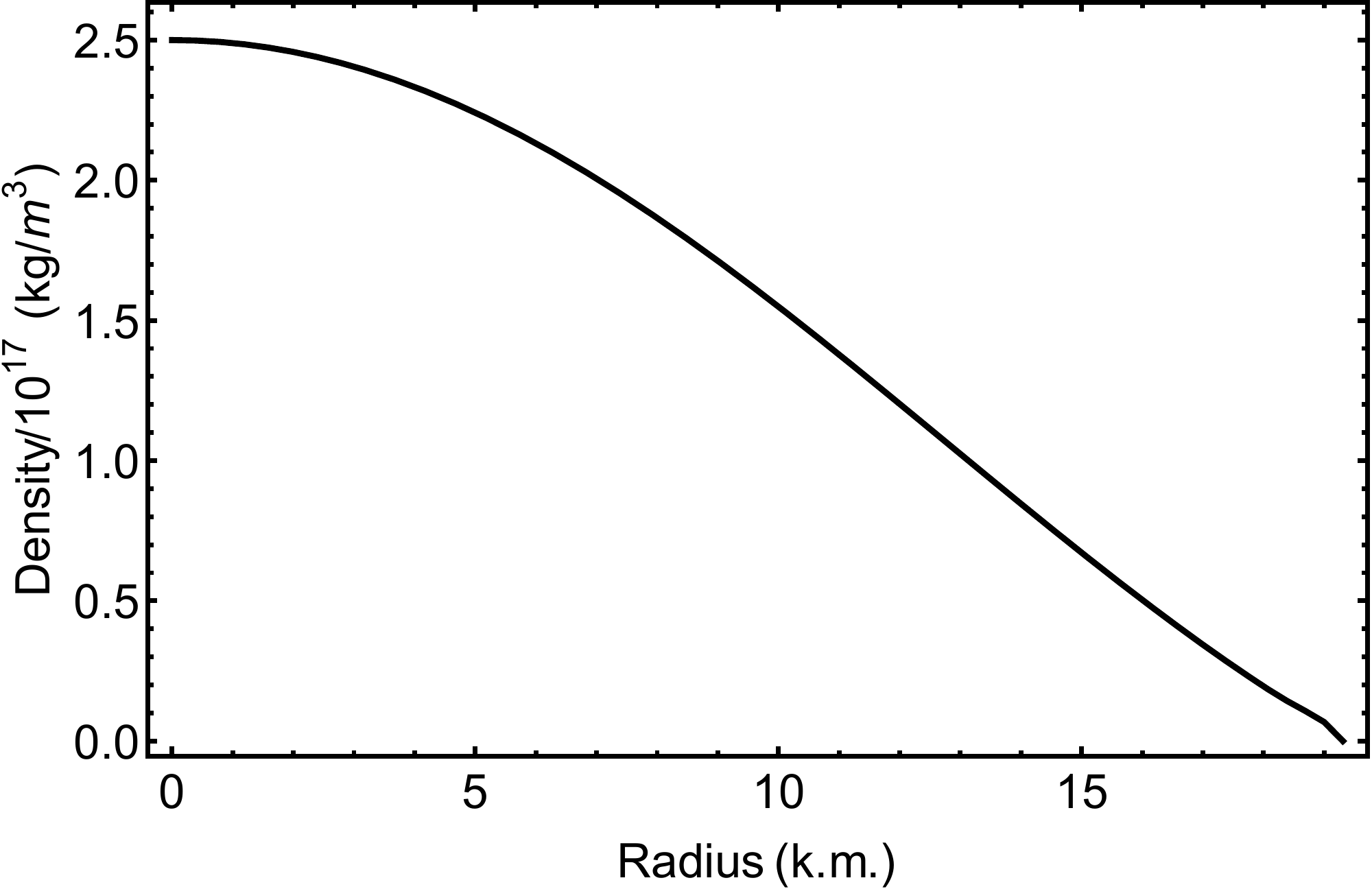} & 
				\includegraphics[width=7cm,height=5cm]{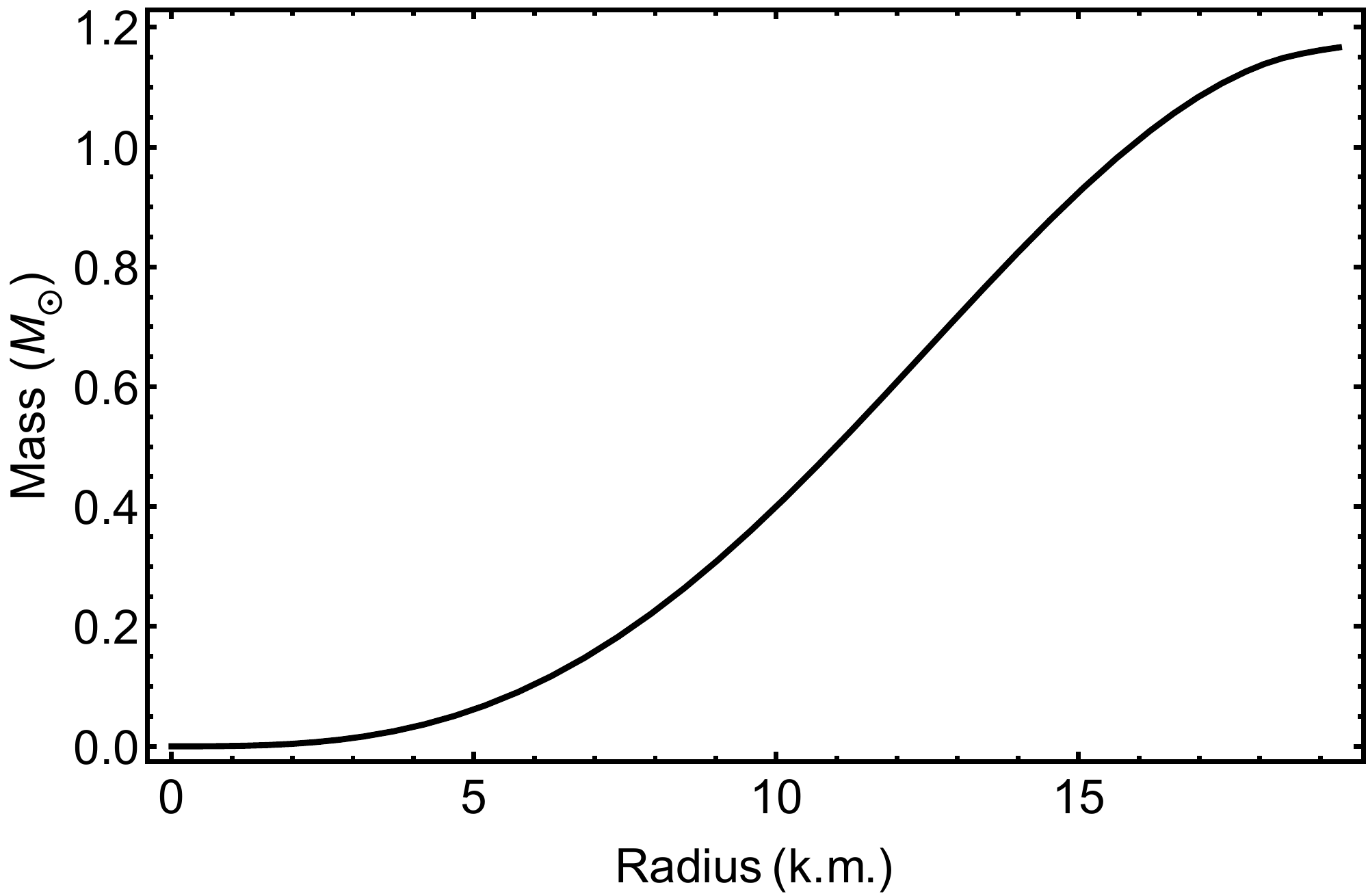}\tabularnewline
				(a) & (b)\tabularnewline
			\end{tabular}
		\end{figure}

		The TOV equation relates the pressure ($P$) with the density ($\rho$) of a spherically compact symmetric star. The pressure $P$, appearing in the above equation, is related to the density ($\rho$) of the star through a polytropic relation, $P=K\rho^{\gamma}$, where $K$ is a constant and $\gamma$ is the polytropic exponent given by, $\gamma=1+\frac{1}{n}$, $n$ being the polytropic index. The variation of mass ($M$) as function of radial distance ($r$) is given by the classical relation,
		\begin{equation}
			\frac{dM}{dr}=4\pi r^{2}\rho.\label{eq:TOV mass}
		\end{equation}
		and describes hydrostatic equilibrium inside the star \cite{shapiro}.
		The mass distribution profile and the density profile are evaluated by solving Eqs.~(\ref{eq:TOV}) and (\ref{eq:TOV mass}) numerically in keeping with suitable parameters. These variations are shown in the plots of Fig.~\ref{fig:profile_of_ns} for an adopted neutron star core density of $2.5\times 10^{17}\:\rm{kg/m}^{3}$. In Fig.~\ref{fig:profile_of_ns}a we show how the density of a neutron star falls off with the increase of radial distance ($r$) from the center of its core, while in Fig.~\ref{fig:profile_of_ns}b the variation of mass of the star enclosed within the radial distance $r$ is plotted. For example, for $r=15$ km the mass enclosed is around $0.95\,M_{\odot}$ while mass within $8\, km$ radius from the center of the neutron star is $0.2\,M_{\odot}$ for this particular example.

		In the present work, a highly compact star (core density $=2.5\times10^{17}\:\rm{kg/m^3}$) is chosen as the lensing star. As mentioned earlier the rays corresponding to the gravitational waves are obtained by solving the equations of null-geodesic (Eqs.~\ref{eq: r_eq} and \ref{eq:theta eq}). The mass $M$ inside the star is obtained from the mass profile, evaluated by solving the TOV equations (Eqs.~\ref{eq:TOV}, \ref{eq:TOV mass}) numerically while outside the star the total mass $\mathcal{M}$ of the star is considered.

		The gravitational waves emanated by the binary mergers modify the space-time in periodic manner during its propagation. So, in order to evaluate its intensity at any particular point, its time variation needs to be taken into account. According the current observational data \cite{GW150914,GW170814,GW170817}, the frequencies of the gravitational waves are in the range of $100\,nHz-1000\,Hz$. Consequently, the maximum phase shift between the waves can be neglected ($\lesssim\pi/50$) near the focused region in the above-mentioned case. So, for the sake of simplicity, the time dependent term is set to unity in our calculation. The effective intensity of the gravitational waves at any point is evaluated in terms of dimensionless intensity ($\mathcal{I}$) given by,
		\begin{equation}
			\mathcal{I}=\dfrac{I}{I_{abs}},\label{eq:r_int}
		\end{equation}
		\noindent where $I$ represent the measured intensity of the waves at that point and $I_{abs}$ would be the intensity if the lensing star is absent.
		
\begin{figure}
			\centering{}
			\caption{\label{fig:gw_trj}Trajectories of the gravitational waves affected by gravitational lensing in presence of the compact star}
			\includegraphics[width=10cm]{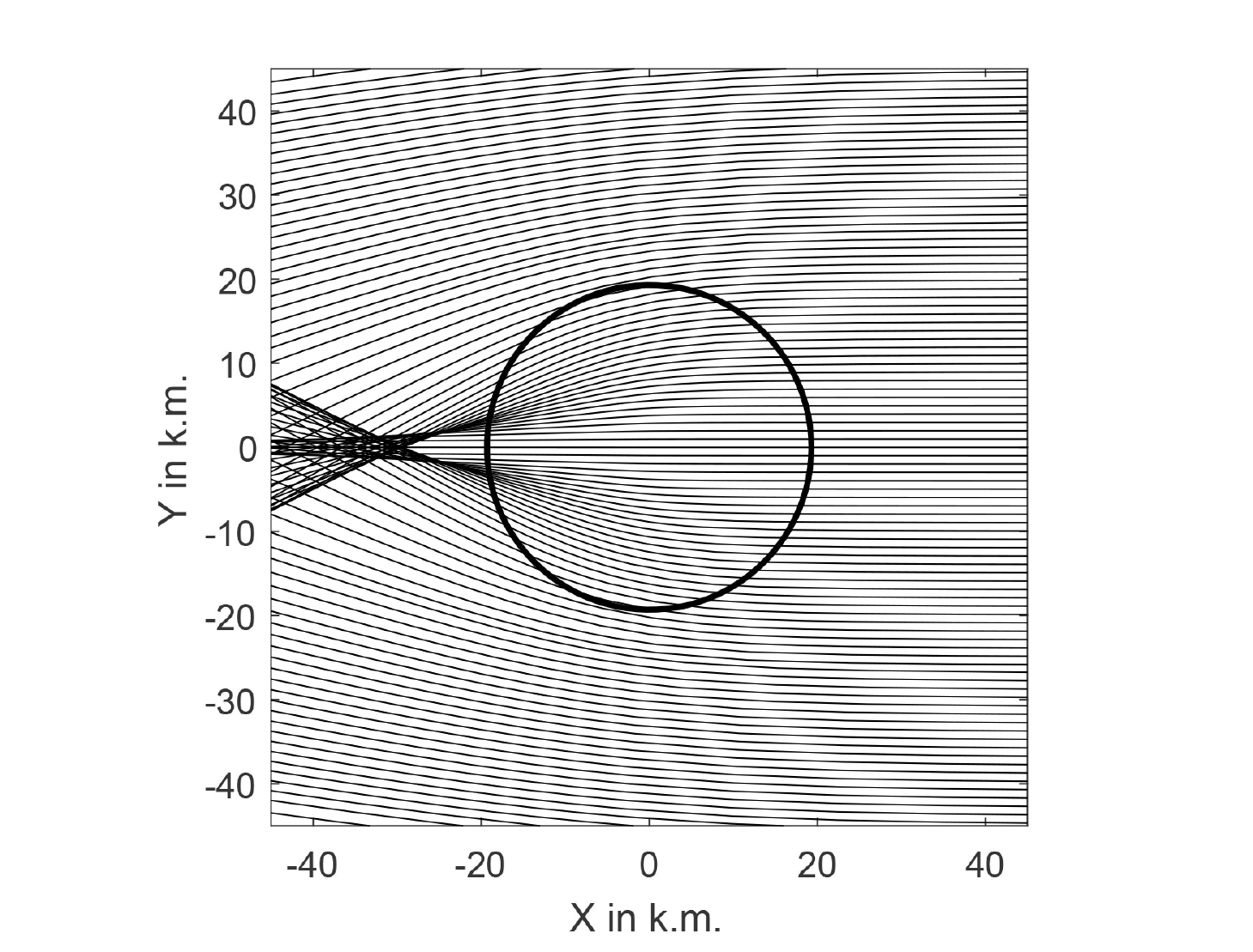}
		\end{figure}

		In Fig.~\ref{fig:gw_trj} we furnish a plot for the trajectories of the gravitational waves in X-Y plane in presence of the chosen compact star to show the lensing effects thus produced.
		It can be seen from Fig.~\ref{fig:gw_trj} that the gravitational waves are affected differently in regions inside the star compared to the way they are affected in the vicinity of the star as the propagating gravitational waves experience qualitatively different space-time distortions inside the star in contrast to its immediate vicinity outside. As a result, these gravitational waves produce dissimilar lensing effects in the presence of a compact object as shown in Fig.~\ref{fig:gw_trj}.
		
		\begin{figure}
			\centering{}
			\caption{\label{fig:GW-intensity-contour}Gravitational waves intensity contour plots as observed from different distances from the star. (a) $x=23$km, (b) $x=25$km, (c) $x=27$km, (d) $x=28$km, (e) $x=29$km, (f) $x=29.5$km, (g) $x=32$km, (h) $x=33$km, (i) $x=48$km}
			\begin{tabular}{cc}
			\begin{tabular}{ccc}
				\includegraphics[width=5cm]{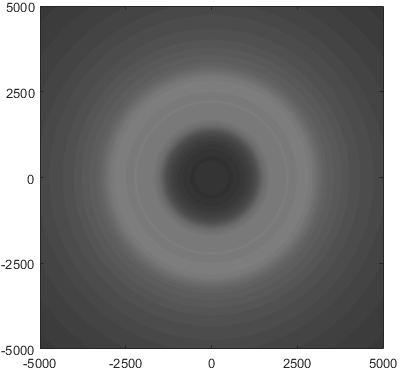} & \includegraphics[width=5cm]{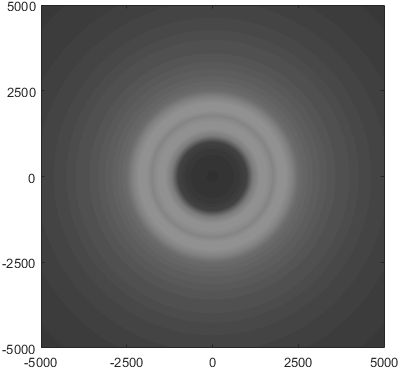} & \includegraphics[width=5cm]{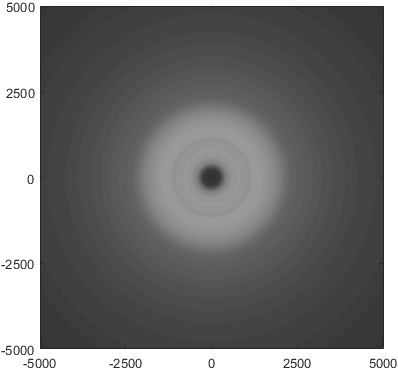}\tabularnewline
				(a) & (b) & (c)\tabularnewline 
				\includegraphics[width=5cm]{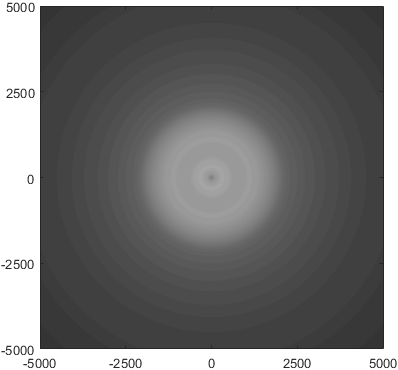} & \includegraphics[width=5cm]{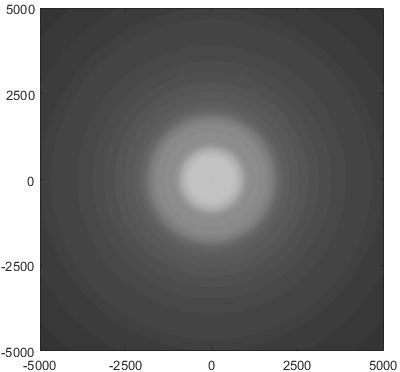} & \includegraphics[width=5cm]{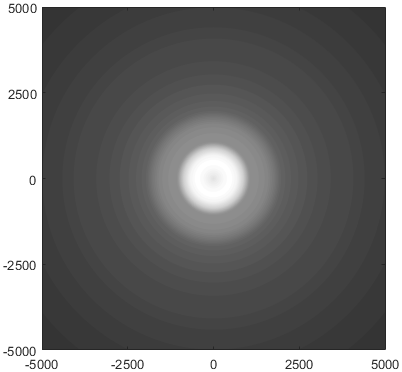}\tabularnewline
				(d) & (e) & (f)\tabularnewline
				\includegraphics[width=5cm]{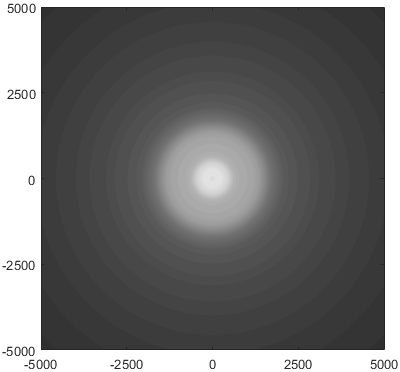} & \includegraphics[width=5cm]{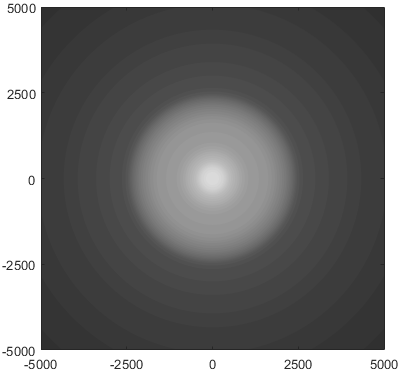} & \includegraphics[width=5cm]{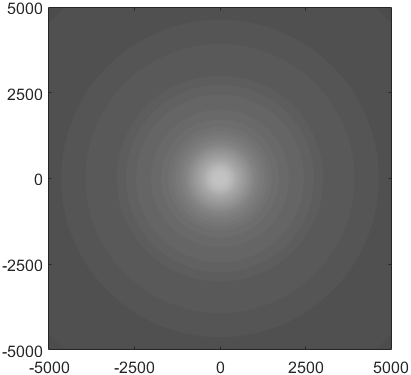}\tabularnewline 
				(g) & (h) & (i)\tabularnewline
			\end{tabular} &
			\begin{minipage}[c]{1.5cm}
				\includegraphics[width=1.5cm,height=10cm]{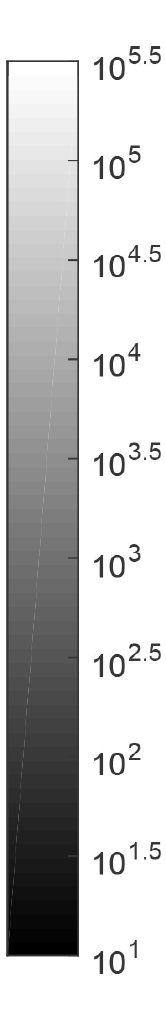}
			\end{minipage}\tabularnewline
			\end{tabular}
		\end{figure}

		While the waves propagate away from the lensing star, fallout in the intensity distribution starts appearing gradually.
		The intensity profiles of the waves in these zones are depicted in Fig.~\ref{fig:GW-intensity-contour}.
		For instance, in the present example, at a distance $x=20$ km from the center of the star the manifestation of lensing effect is confined in a ring-shaped region in space (Fig.~\ref{fig:GW-intensity-contour}). The extent of the ringed shaped region gradually shrinks down as the distance $x$ increases (Fig.~\ref{fig:GW-intensity-contour}) with gradual increase in intensity. Eventually the intensity of the luminous portion of the ring attains maximum at $x\sim25$ km. The ring continuous to shrink afterwards and at a distance $x=29.5$ km we get a sharply focused point with maximum intensity.
		The gravitational waves are focused on a range of points located on the propagation axis (principle axis in this case), rather than a single point (Fig.~\ref{fig:gw_trj}). This phenomenon is comparable to aberration observed in thick optical lens.
		
\begin{figure}
			\centering{}
			\caption{\label{fig:int_max x}Variation of Maximum intensity with distance ($x$) from the lensing star. The distance corresponding to the peak P is the characteristic distance for the lensing star.}
			\includegraphics[width=10cm,height=6cm]{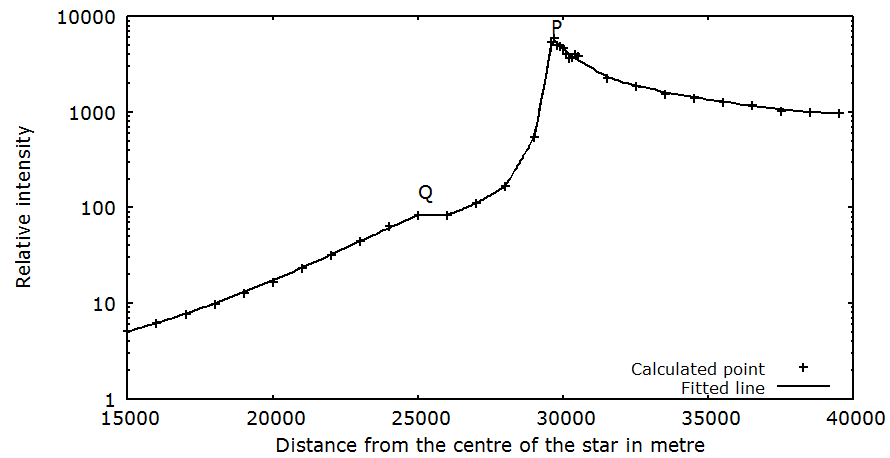}
		\end{figure}

		The variation of the relative intensities due to lensing of gravitational waves with the distance from the center of the star are plotted in Fig.~\ref{fig:int_max x} for the demonstrative example considered in this work. From Fig.~\ref{fig:int_max x} one observes that the peak relative intensity at a distance $x=29.5$ km is $\sim10^{4}$ (point P) which corresponds to the best focused point shown in Fig.~\ref{fig:GW-intensity-contour}f.
		At this distance the gravitational waves are focused to a point. Fig.~\ref{fig:int_max x} provides another smaller peak at Q ($\mathcal{I}\sim100$) at a distance $x\sim25$ km from the center of the star. It also depicts the variation of intensity inside the extended zone of focus shown in Fig.~\ref{fig:GW-intensity-contour}b. 
		The lensing power of the star can be characterized by the distance to the best focused point (point P) which would called the characteristic distance $\mathcal{D}$ henceforth. The characteristic distance $\mathcal{D}$ of the star is analogous to the focal length of an optical lens. 
		As the observer goes beyond $\mathcal{D}$ the maximum intensity starts falling again almost exponentially with the distance $x$ and the central focal spot completely disappear at large distance ($x\rightarrow\infty$) eventually.

\begin{figure}
			\centering{}
			\caption{\label{fig:int_max x-1} Characteristic distance $\mathcal{D}$ vs mass of the star}
			\includegraphics[height=7cm]{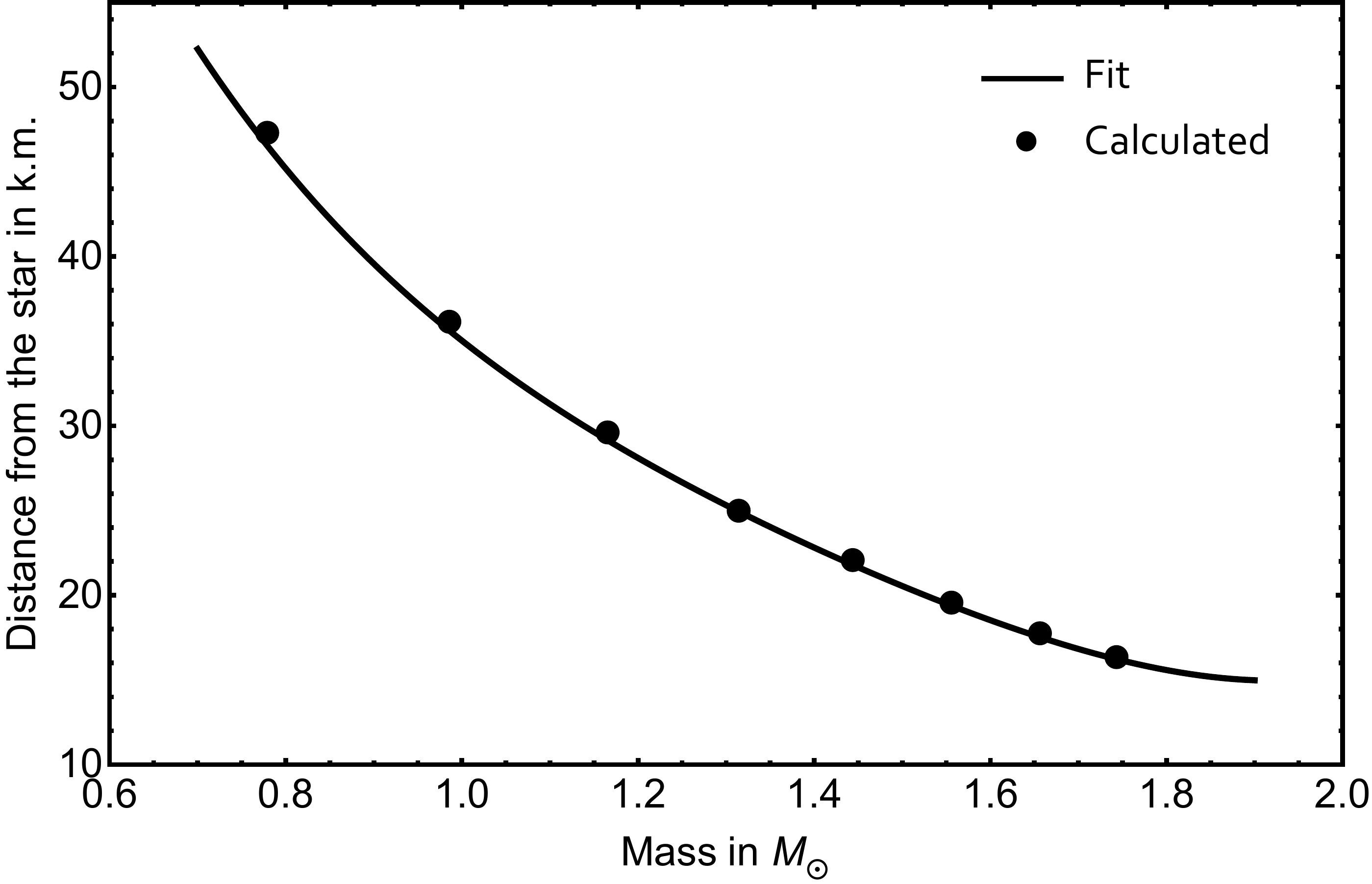}
		\end{figure}

		For the present example $\mathcal{D}=29.6$ km. The dependence of $\mathcal{D}$ on the masses of the lensing star have been found by repeating the same calculation with other stars of different masses. In Fig.~\ref{fig:int_max x-1} the variation of $\mathcal{D}$ with the mass of the lensing stars is shown, from which it is evident that, $\mathcal{D}$ decreases as the mass of the lensing star increases. This is due to the fact that more massive the star is more deformation it would cause to its adjacent space-time that eventually increases the focusing effect. The characteristic distance $\mathcal{D}$ also changes accordingly. To examine whether there is a systematic formula that matches the dependence, we have fitted our numerical results shown in Fig.~\ref{fig:int_max x-1} with an assumed empirical relation for $\mathcal{D}$ given by,
		\begin{equation}
			\mathcal{D}=\dfrac{2GM_{\odot}}{c^2}\dfrac{9r^2}{c}\bigg/\left(\dfrac{\mathcal{M}}{M_{\odot}}\right).
			\label{eq:d_mass}
		\end{equation}
		In Eq.~\ref{eq:d_mass}, $r$ denotes the radius of the star of mass $\mathcal{M}$ which is obtained by solving the TOV equation. From the fitted graph (Fig.~\ref{fig:int_max x-1}) one sees that the Eq.~\ref{eq:d_mass} is a good fit with ($\chi^2=0.0019$ \footnote{$\chi^{2}=\sum\left(\frac{N_{D}-F_{D}}{N_{D}}\right)^{2}$, where $N_{D}$ is the numerically obtained value and $F_{D}$ is the fitted value.}) our numerical results.		 
		
		In this work, we have drawn an analogy between the lensing effects in gravitational waves by massive compact star and that of ordinary light by thick optical lens. For the sake of simplicity the waves are presumed to be emitted by highly red-shifted source. As a consequence, the incident waves can be considered to be plane waves, rather than spherical in nature.
		
		In the present example, the propagating waves after suffering the lensing effect in presence of a massive star tends to be localized or focused on the other side of the lensing star instead of being focused at a single point. This indicates the aberration effects in case of gravitational lensing analogous to the optical aberration related to a thick optical lens. The relative intensity of the gravitational waves after they are subjected to the lensing effect, is found to be maximized at a distance $\mathcal{D}$, termed as characteristic distance from the center of the star. The magnitude of this distance $\mathcal{D}$ depends on the mass of the lensing star as well as its core density and this variation seems to follow an empirical relation given in Eq.~\ref{eq:d_mass}. 
				
		In summary, we have studied the lensing effects of gravitational waves by massive lensing stars. We elaborate the similarities of the gravitational lensing phenomenon by a compact massive star with the optical lensing of light waves in the presence of a thick convex lens. Our studies demonstrate that the phenomenon of such lensing of GWs and their consequent convergence effects with enhanced intensities would help in interpreting gravitational wave signatures from deep space.
	\section*{Acknowledgment}	
		
		Two of the authors (S.B. and A.H.) wish to acknowledge the support received from St. Xavier's College, kolkata Central Research Facility. A.H. also acknowledges the University Grant Commission (UGC) of the Government of India, for providing financial support, in the form of UGC-CSIR NET-JRF.

	\bibliographystyle{aipauth4-1}
	\bibliography{intensification_gw}

\end{document}